\documentclass[english,preprint]{revtex4-1}
\usepackage{libertineRoman}
\usepackage[T1]{fontenc}
\usepackage[latin9]{inputenc}
\usepackage{color}
\usepackage{babel}
\usepackage{units}
\usepackage{amsmath}
\usepackage{amsthm}
\usepackage{amssymb}
\usepackage{graphicx}
\usepackage[unicode=true,pdfusetitle,
 bookmarks=true,bookmarksnumbered=false,bookmarksopen=false,
 breaklinks=false,pdfborder={0 0 1},backref=false,colorlinks=true]
 {hyperref}
\hypersetup{
 urlcolor=blue, citecolor=goodgreen, linkcolor=goodgreen}
\usepackage{breakurl}

\makeatletter
\definecolor{goodgreen}{rgb}{0, 0.4, 0.8}

\makeatother

\begin{document}

\title{Quantum change point and entanglement distillation }

\author{Abhishek Banerjee}
\email{banerjee.abhishek.official@gmail.com }

\author{Pratapaditya Bej}
\email{pratap6906@gmail.com }

\author{Somshubhro Bandyopadhyay}
\email{som@jcbose.ac.in, som.s.bandyopadhyay@gmail.com}

\affiliation{Department of Physical Sciences, Bose Institute, Bidhannagar, Kolkata
700091, India}
\begin{abstract}
In a quantum change point problem, a source emitting particles in
a fixed quantum state (default) switches to a different state at some
stage, and the objective is to identify when the change happened by
measuring a sequence of particles emitted from such a source. Motivated
by entanglement-sharing protocols in quantum information, we study
this problem within the paradigm of local operations and classical
communication (LOCC). Here, we consider a source that emits entangled
pairs in a default state, but starts producing another entangled state
(mutation) at a later stage. Then, a sequence of entangled pairs prepared
from such a source and shared between distant observers cannot be
used for quantum information processing tasks as the identity of each
entangled pair remains unknown. We show that identifying the change
point using LOCC leads to the distillation of free entangled pairs.
In particular, if the default and the mutation are mutually orthogonal,
there exists an efficient LOCC protocol that identifies the change
point without fail and distills a sufficiently large number of pairs.
However, if they are nonorthogonal, there is a probability of failure.
In this case, we compute the number of entangled pairs that may be
obtained on average. We also consider a relaxation of the two-state
problem where the mutation is not known \emph{a priori,} but instead
belongs to a known set. Here we show that local distinguishability
plays a crucial role: if the default and the possible mutations are
locally distinguishable, the problem reduces to the two-state problem
with orthogonal states, but if not, one may still identify the mutation,
the change point, and distill entanglement, as we illustrate with
a concrete example. 
\end{abstract}
\maketitle
The change point detection in statistical analysis seeks to identify
variations in probability distributions in stochastic processes or
time series data with a range of applications in diverse fields. This
problem has been recently formulated in the quantum domain \citep{Aki-Hayashi-2011,Sentis+-2016}
and has received particular attention \citep{Sentis+2017,Sentis+2018,Fanizza+2023}.
Here, one considers a quantum source that emits particles in a fixed
(default) state $\left|\psi\right\rangle $. The source, however,
changes its character (mutation) at some point during emission and
starts producing particles in another state $\left|\phi\right\rangle $
nonorthogonal to $\left|\psi\right\rangle $ \citep{Note-1}. Now,
given a sequence of $n$ particles prepared from such a source, the
objective is to determine at which point the change took place, assuming
that every point of the given sequence is equally likely to be the
change point. 

The problem can be formulated as a state discrimination task \citep{Sentis+-2016}.
Since the state of a sequence assumed to have the change point at
the $k^{\text{th}}$ position can be written as
\begin{flalign}
\left|\xi_{k}\right\rangle  & =\left|\psi\right\rangle ^{\otimes k-1}\left|\phi\right\rangle ^{\otimes n-k+1},\label{xi-k-1}
\end{flalign}
the problem of detecting the change point boils down to the task of
distinguishing the states $\left|\xi_{1}\right\rangle ,\dots,\left|\xi_{n}\right\rangle $.
This can be solved using the minimum-error approach in the asymptotic
limit \citep{Sentis+-2016} or the unambiguous discrimination approach
for any $n$ \citep{Sentis+2017,Sentis+2018}. Note that the former
determines the maximum probability of identifying the change point
correctly but allows for errors, whereas the latter determines the
change point without error but with a nonzero probability. Very recently,
how quickly the change point can be detected has been investigated
\citep{Fanizza+2023}. 

The present paper considers the quantum change point (QCP) problem
in the paradigmatic \textquotedblleft distant laboratory\textquotedblright{}
scenario in quantum information theory. The paradigm involves a composite
system, parts of which are shared among several remote observers who
are restricted to acting on respective subsystems and communicating
classically. Quantum operations realized this way are said to belong
to the local operations and classical communication (LOCC) class \citep{LOCC}.
Many fundamental questions, especially those concerning quantum nonlocality
and quantum resources, such as entanglement, are studied within the
framework of LOCC.

Our motivation stems from the working of entanglement-sharing protocols.
These protocols are central to quantum information because shared
entanglement is a critical resource for tasks that are either impossible
or sub-optimal otherwise. In a typical entanglement-sharing protocol,
a source prepares entangled pairs in a known state, and quantum channels
distribute these pairs to remote observers \citep{Horodecki+2008}.
The quantum channels, in general, are noisy, resulting in mixed (noisy)
entangled states. However, this can be addressed by entanglement distillation
protocols to obtain fewer entangled states of high purity from a collection
of such mixed states \citep{Bennett+1996-1,Bennett+1996}. 

The QCP problem fits naturally in an entanglement-sharing setup once
we recognize the source could be faulty, i.e., may have mutated at
some stage and started producing a different entangled state. This,
however, has a nontrivial implication. To see this, consider a two-party
setup and that the entangled pairs from a possibly faulty source are
shared between two remote observers, Alice and Bob, where we assume
the quantum channels are noiseless. First note that had the quantum
source not malfunctioned, Alice and Bob would share a sequence of
pure entangled states that they could use for quantum information
processing tasks. But now, with the possibility that the source may
have switched to another state during emission, they can no longer
be sure of the identities of the pairs they hold. Since they do not
know at which point the switch happened (if at all), they will effectively
share a mixed state that may no longer be useful. This is, therefore,
purely a consequence of the occurrence of a change point. Thus, the
QCP problem in an LOCC setting goes beyond answering the question
of just detecting the change point. 

Here, we will show that identifying the change point using LOCC leads
to the \textquotedblleft distillation\textquotedblright{} of free
entangled pairs. First, we will consider the two-state problem, defined
along the lines of the standard QCP formulation but now with entangled
states. We will show that if the concerned states are mutually orthogonal,
there exists an efficient LOCC protocol that not only identifies the
change point without fail but also distills a sufficiently large number
of free entangled pairs. For nonorthogonal states, a general analysis
seems to be hard. Nevertheless, we derive a recurrence relation to
compute the average entanglement that may be distilled by LOCC. 

One could now argue that in a QCP formulation, the assumption that
the mutation (switched state) is known beforehand is rigid and should
be relaxed. One may consider, for example, the other extreme where
the mutation is unknown. However, identifying the change point without
error would then be impossible. In this paper, we will consider an
intermediate scenario where we suppose the mutation is not known \emph{a
priori,} but is a member of a known set of entangled states. For simplicity,
we will suppose the default state and all possible mutations are mutually
orthogonal. We will show that if the states (default plus mutations)
are LOCC distinguishable, the problem reduces to the two-state problem
with orthogonal states. However, if they are not, while entanglement
distillation is always possible and mostly efficient, detecting the
change point and the mutation may not be possible. This is explicitly
discussed by considering the Bell basis, where one of the Bell states
is taken to be the default and the rest as mutations. 

Let us consider a source emitting entangled qudit (a $d$-dimensional
quantum system) pairs in a fixed state $\left|\Psi\right\rangle $
that are being shared between Alice and Bob. At some stage, the source
switches to a different entangled state $\left|\Phi\right\rangle $
that, in general, is nonorthogonal to $\left|\Psi\right\rangle $.
Then, given a sequence of $n$ ($n$ is large) entangled pairs prepared
from such a source, the objective is to identify at which point (from
which pair onward) the change happened using LOCC, where we assume
that the position of every pair of a given sequence is equally likely
to be the change point including the possibility that no change occurs.
Note that we have implicitly assumed that the involved quantum channels
are noiseless (this will hold throughout the paper). This is made
for clarity, to single out the implications of a change point. 

The state of a sequence with the change point at the $k^{\text{th}}$
position for $k\in\left\{ 1,\dots,n\right\} $ is of the form
\begin{alignat}{1}
\left|\Gamma_{k}\right\rangle  & =\left|\Psi\right\rangle ^{\otimes k-1}\left|\Phi\right\rangle ^{\otimes n-k+1}.\label{Gamma-k}
\end{alignat}
If no change occurs, the state is $\left|\Psi\right\rangle ^{\otimes n}$,
obtained for $k=n+1$ in \eqref{Gamma-k}. The change point, if any,
can therefore be identified by distinguishing the states $\left\{ \left|\Gamma_{k}\right\rangle \right\} _{k=1}^{n+1}$
by LOCC, where each occurs with probability $\nicefrac{1}{\left(n+1\right)}$. 

Note that Alice and Bob actually share one of $\left|\Gamma_{1}\right\rangle ,\dots,\left|\Gamma_{n+1}\right\rangle $,
each representing a sequence of $n$ entangled pairs but none can
be used as their identities remain unknown. This implies they share
the mixed state 
\begin{flalign}
\rho_{\Gamma} & =\frac{1}{n+1}\sum_{k=1}^{n+1}\left|\Gamma_{k}\right\rangle \left\langle \Gamma_{k}\right|\label{rho-Gamma}
\end{flalign}
which may not be of use for quantum information tasks. 

We now discuss how detecting the change point leads to entanglement
distillation from the mixed state \eqref{rho-Gamma}. The argument
is simple. Suppose someone informs Alice and Bob that the change happened
at $k=k_{c}$. This information singles out the sequence $\left|\Gamma_{k_{c}}\right\rangle $
and they can now conclude that all pairs before $k_{c}$ are in $\left|\Psi\right\rangle $
and the rest are in $\left|\Phi\right\rangle $, which amounts to
the release of free entangled pairs from $\rho_{\Gamma}$, and hence
distillation \citep{Note-2}. Therefore, if they can identify the
change point, without error, by LOCC but without consuming all pairs,
they will clearly have useful entanglement. So, the objective here
is not just to identify the change point (without error), but to do
so in a way that will keep the number of entangled pairs consumed
in the process at a minimum (see Fig.$\,$1). 

\begin{figure}
\includegraphics[scale=0.35]{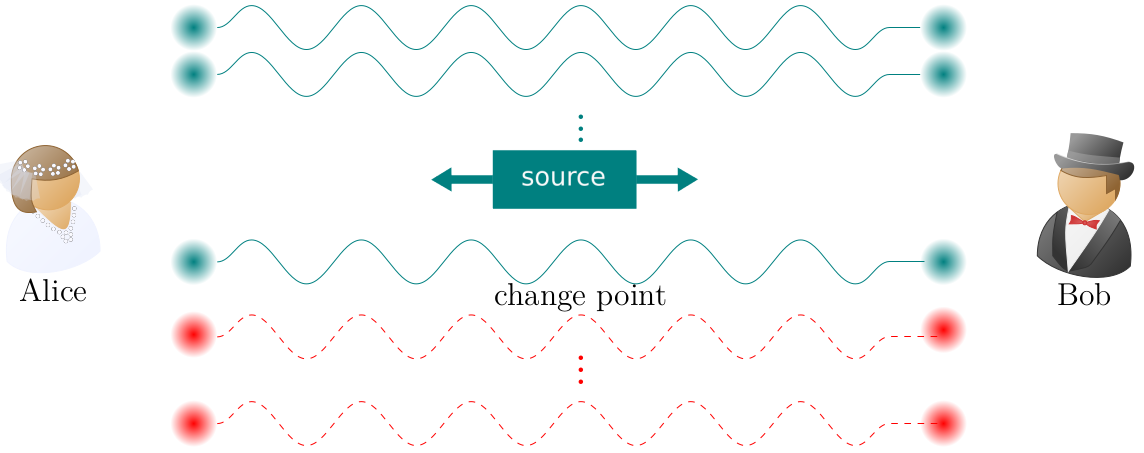}\caption{LOCC quantum change point problem: Two remote observers, Alice and
Bob, share a sequence of entangled pairs emitted by a possibly faulty
source. The green pairs are in the default state $\left|\Psi\right\rangle $
and the red ones are in the mutated state $\left|\Phi\right\rangle $.
As explained in the text, the task is to identify the change point
and distill free entangled pairs using LOCC. }
\end{figure}

The above discussion suggests that Alice and Bob perform local measurements
on individual entangled pairs until the change point is identified.
However, for this approach to succeed, the LOCC measurements on a
pair must reveal its identity with certainty or, at the very least,
with nonzero probability. Fortunately, this is possible, thanks to
the following results from LOCC state discrimination: the state of
a composite system, known to be in one of two pure states, can be
correctly determined by LOCC with nonzero probability \citep{Virmani+-2001},
and this probability is one iff the states are orthogonal \citep{Walgate-2000}.
Thus, the state of any pair of a given sequence can be determined
by LOCC with nonzero probability if $\left\langle \Psi\vert\Phi\right\rangle \neq0$
or with certainty if $\left\langle \Psi\vert\Phi\right\rangle =0$,
although in the process the state loses entanglement \citep{Virmani+-2001,Walgate-2000}. 

First, consider the case $\left\langle \Psi\vert\Phi\right\rangle =0$.
To avoid confusion, denote the mutation by $\left|\Psi^{\perp}\right\rangle $
and the change point, if the change occurs, by $k_{c}$. It follows
that unless $k_{c}=1$ or no change occurs, the change point will
be detected once Alice and Bob find adjacent pairs in states $\left|\Psi\right\rangle $
and $\left|\Psi^{\perp}\right\rangle $, respectively, the position
of the latter being the change point; if $k_{c}=1$ or no change occurs,
then, for the former, it will be detected once the first pair is identified
as $\left|\Psi^{\perp}\right\rangle $, whereas for the latter, the
last pair as $\left|\Psi\right\rangle $. The question is whether
this could be done efficiently by LOCC, i.e. without consuming a lot
of pairs. We now show that this is indeed possible. 

\begin{figure}
\includegraphics[scale=0.35]{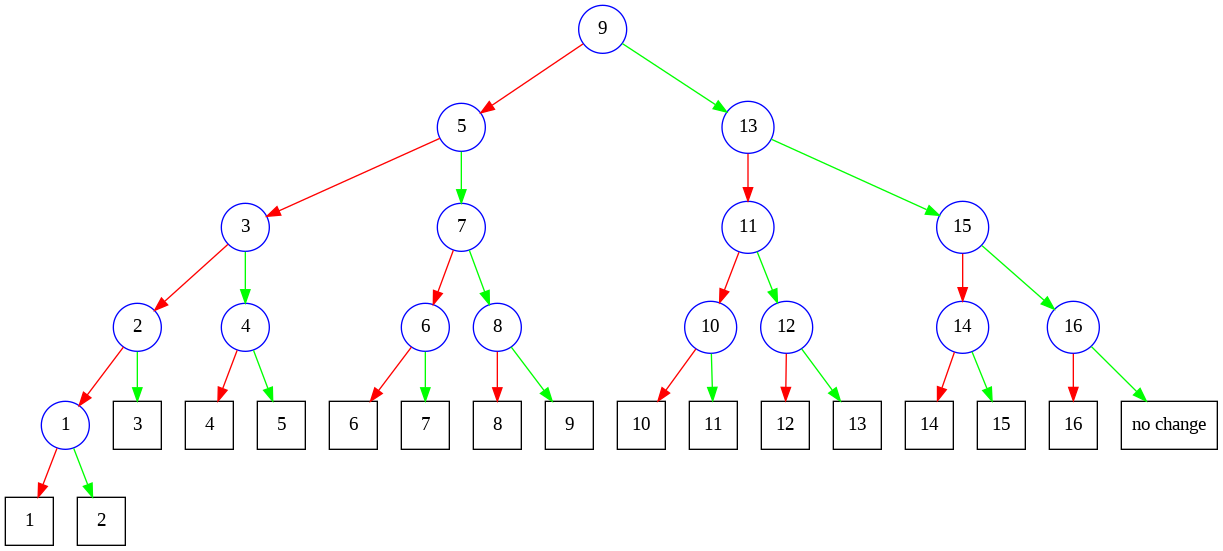}

\caption{The two-state protocol with orthogonal states for $n=16$. The numbers
inside the squares indicate the change points, and those inside the
circles indicate the positions of the entangled pairs on which measurements
are performed. The green and red arrows indicate the outcomes $\Psi$
and $\Psi^{\perp}$, respectively. The tree depicts all possible measurement
sequences detecting the change points, including when no change occurs.
Here, the worst-case scenario requires five measurements.}
\end{figure}

The LOCC protocol is inspired by the binary search algorithm \citep{Knuth}
and is explained in the Appendix. Briefly it works as follows. First,
the state at $i_{\text{mid}}$, the midpoint of the sequence, is determined:
(a) Outcome $\Psi$ implies $i_{\text{mid}}<k_{c}\leq n$ and all
pairs $\left\{ i\right\} $ for $i<i_{\text{mid}}$ are in $\left|\Psi\right\rangle $;
(b) however, outcome $\Psi^{\perp}$ implies $1\leq k_{c}\leq i_{\text{mid}}$
and all pairs $\left\{ i\right\} $ for $\left(i_{\text{mid}}+1\right)\leq i\leq n$
are in $\left|\Psi^{\perp}\right\rangle $. Thus outcome $\Psi$ frees
up or distills $\left(k_{\text{mid}}-1\right)$ pairs in $\left|\Psi\right\rangle $,
whereas $\Psi^{\perp}$ distills $\left(n-k_{\text{mid}}\right)$
pairs in $\left|\Psi^{\perp}\right\rangle $. Now, depending on the
outcome ($\Psi$ or $\Psi^{\perp}$) in the first step, the protocol
continues with the pairs either on the right of $i_{\text{mid}}$
(for $\Psi$) or on the left of $i_{\text{mid}}$ (for $\Psi^{\perp}$).
The procedure is repeated till $k_{c}$ is detected or no change conclusion
is reached (Fig.$\,$2 illustrates it for $n=16$). The protocol is
guaranteed to succeed and requires either $\left\lfloor \log_{2}\left(n+1\right)\right\rfloor $
or $\left(\left\lfloor \log_{2}n\right\rfloor +1\right)$ measurements
(see Appendix A) . Since every LOCC measurement consumes a pair, the
number of residual (distilled) entangled pairs is either $\left(n-\left\lfloor \log_{2}\left(n+1\right)\right\rfloor \right)$
or $\left(n-\log_{2}\left\lfloor n\right\rfloor -1\right)$. Note
that the residual pairs before $k_{c}$ are in $\left|\Psi\right\rangle $
and those after $k_{c}$ are in $\left|\Psi^{\perp}\right\rangle $;
however, all residual pairs will be in $\left|\Psi^{\perp}\right\rangle $
if $k_{c}=1$ or $\left|\Psi\right\rangle $ if there is no change. 

Thus we have shown that one can identify the change point and also
distill a reasonably large number ($\approx n-\log_{2}n$) of free
entangled pairs from $\rho_{\Gamma}$ {[}\eqref{rho-Gamma}{]} if
the default and the mutation are mutually orthogonal. This analysis
holds for any choice of $\left|\Psi\right\rangle $ and $\left|\Psi^{\perp}\right\rangle $. 

We now come to the general two-state problem where $0<\left\langle \Psi\vert\Phi\right\rangle <1$.
As noted earlier, $\left|\Psi\right\rangle $ and $\left|\Phi\right\rangle $
can be unambiguously distinguished by LOCC with a nonzero probability
\citep{Virmani+-2001}, where there are three possible outcomes of
a measurement: two are conclusive, detecting $\Psi$ and $\Phi$,
respectively, but the third is inconclusive, in which case we do not
get to know about the identity. The LOCC protocol is the same as before,
with an obvious modification: proceed the same way if the outcome
is conclusive, else discard the pair and work with the rest (since
we are discarding, one may never identify the change point). The best
case is clearly when every measurement is successful, whereas the
worst case is where every measurement is inconclusive until we reach
an array of two--then, neither the change point can be identified
nor any pair can be distilled. We will, therefore, consider the average
case scenario and obtain a recursion relation to compute the average
entanglement that may be obtained. 

First note that because of the nature of the possible sequences, the
prior probabilities of $\left|\Psi\right\rangle $ and $\left|\Phi\right\rangle $
associated with a pair depend on the length of the sequence under
consideration. Moreover, the probabilities of the measurement outcomes
depend on the prior probabilities and the inner product \citep{prior-probs}. 

Suppose, at some stage, the sequence consists of $m$ pairs, where
$m\leq n$. Recall that the protocol requires that the measurement
be performed on the pair at the midpoint $i_{\text{mid}}=\left\lfloor \frac{m}{2}\right\rfloor +1$.
Then the prior probabilities are given by $p_{\Psi}\left(m\right)=1-p_{\Phi}\left(m\right)$,
where $p_{\Phi}\left(m\right)=\frac{\left\lfloor \frac{m+1}{2}\right\rfloor }{m+1}$.
Let the probabilities of the conclusive outcomes be $\tilde{p}_{\Psi}\left(m\right)$
and $\tilde{p}_{\Phi}\left(m\right)$, and, therefore, the failure
probability is $p_{f}\left(m\right)=1-\tilde{p}_{\Psi}\left(m\right)-\tilde{p}_{\Phi}\left(m\right)$.
Then, (a) an inconclusive outcome leaves a sequence of $m-1$ pairs
with probability $p_{f}\left(m\right)$, whereas (b) a conclusive
outcome $\Psi$ ($\Phi$) leaves a sequence of length $\left\lfloor \frac{m-1}{2}\right\rfloor $
( $\left\lfloor \frac{m}{2}\right\rfloor $) with probability $\tilde{p}_{\Psi}\left(m\right)$
{[}$\tilde{p}_{\Phi}\left(m\right)${]} on the right (left) of $i_{\text{mid}}$.
Let $\overline{N_{m}}$ denote the average number of entangled pairs
used up during the protocol starting from a sequence of length $m$.
The following recursion holds: 
\begin{alignat}{1}
\overline{N}_{m} & =1+p_{f}\left(m\right)\overline{N}_{m-1}+\tilde{p}_{\Psi}\left(m\right)\overline{N}_{\left\lfloor \frac{m-1}{2}\right\rfloor }+\tilde{p}_{\Phi}\left(m\right)\overline{N}_{\left\lfloor \frac{m}{2}\right\rfloor }\label{recursion}
\end{alignat}
with the initial condition $\overline{N_{0}}=0$. Therefore, on average,
the number of entangled pairs distilled from a sequence of $n$ entangled
pairs is $n-\overline{N_{n}}$. 

We now come to the relaxation of the two-state problem. Here, the
source switches to one of $m\geq2$ mutually orthogonal entangled
states $\left|\Phi_{1}\right\rangle ,\dots,\left|\Phi_{m}\right\rangle $
with equal probability, but we do not know which. However, once the
switch happens, the mutation remains unchanged through subsequent
emissions. We further assume the mutations are orthogonal to the default
state $\left|\Psi\right\rangle $. 

The state of a sequence having the change point at the $k^{\text{th}}$
position is given by
\begin{alignat}{1}
\left|\omega_{k,i}\right\rangle  & =\left|\Psi\right\rangle ^{\otimes k-1}\left|\Phi_{i}\right\rangle ^{n-k+1}\label{eomega-k-i}
\end{alignat}
for some $\left|\Phi_{i}\right\rangle $; the ``no change'' scenario,
as before, corresponds to $k=n+1$ with the sequence represented by
$\left|\omega_{n+1}\right\rangle $. Since $\left|\Phi_{i}\right\rangle $
is unknown, a sequence is described by the density operator
\begin{alignat}{1}
\Omega_{k} & =\frac{1}{m}\sum_{i=1}^{m}\left|\omega_{k,i}\right\rangle \left\langle \omega_{k,i}\right|\label{Omega-k}
\end{alignat}
for $k\in\left[1,n\right]$, and the shared state by
\begin{alignat}{1}
\rho_{\Omega} & =\frac{1}{n+1}\sum_{k=1}^{n+1}\Omega_{k},\label{rho-omega}
\end{alignat}
where $\Omega_{n+1}=\left(\left|\Psi\right\rangle \left\langle \Psi\right|\right)^{\otimes n}$. 

The change point (if any) can be identified by distinguishing between
$\left(nm+1\right)$ orthogonal states $\left\{ \left|\omega_{k,i}\right\rangle ,\left|\omega_{n+1}\right\rangle \right\} $,
$k=1,\dots,n$, $i=1,\dots,m$ by LOCC. The objective, as before,
is to design an efficient LOCC protocol that will identify the change
point and distill entanglement from $\rho_{\Omega}$. Previously,
in the two-state problem with orthogonal states, the LOCC protocol
exploited the fact that the state of any entangled pair can be determined
with certainty. Here, this will require distinguishing between the
states $\left|\Psi\right\rangle ,\left|\Phi_{1}\right\rangle ,\dots,\left|\Phi_{m}\right\rangle $
by LOCC \citep{Peres-Wootters-1991,ben99,walgate-2002,HSSH,Halder+-2019,Ghosh+2001,Watrous-2005,Nathanson-2005}. 

There are now two possibilities: the states are either LOCC distinguishable
or they are not. If they are, the state of any pair can be determined
with certainty by LOCC, and the problem reduces to the two-state problem
with orthogonal states for which we know an efficient LOCC protocol
exists. The nontrivial situation arises when they are not LOCC distinguishable,
which implies the state of a pair cannot be determined using LOCC.
A general analysis of this problem appears difficult, so we consider
an example that captures many essential features. 

Let the default state be $\left|\Psi\right\rangle =1/\sqrt{2}\left(\left|00\right\rangle +\left|11\right\rangle \right)$,
and the possible mutations be $\left|\Phi_{1}\right\rangle =1/\sqrt{2}\left(\left|00\right\rangle -\left|11\right\rangle \right)$,
$\left|\Phi_{2}\right\rangle =1/\sqrt{2}\left(\left|01\right\rangle +\left|10\right\rangle \right)$,
and $\left|\Phi_{3}\right\rangle =1/\sqrt{2}\left(\left|01\right\rangle -\left|10\right\rangle \right)$.
The four Bell states, however, are LOCC indistinguishable \citep{Ghosh+2001};
in particular, given any one of the four Bell states, it is impossible,
even with a nonzero probability, to identify it correctly by LOCC.
So a different protocol is now required. 

First note that even though the state of a pair cannot be determined,
a good deal of information can still be extracted by measuring the
pair in the computational basis $\left\{ \left|ij\right\rangle :i,j=0,1\right\} $:
outcomes $00/11$ or $01/10$ would reveal whether it was one of $\left\{ \left|\Psi\right\rangle ,\left|\Phi_{1}\right\rangle \right\} $
or $\left\{ \left|\Phi_{2}\right\rangle ,\left|\Phi_{3}\right\rangle \right\} $,
respectively. We will now consider a protocol along the lines of the
previous one, but with a mix of LOCC measurements. The protocol begins
by measuring the pair at the midpoint of the sequence $i_{\text{mid}}$
in the computational basis. \\
(a1) If the outcome is $00/11$, then the state was either $\left|\Psi\right\rangle $
or $\left|\Phi_{1}\right\rangle $. Then the preceding pair must also
be in $\left|\Psi\right\rangle $ or $\left|\Phi_{1}\right\rangle $.
Since $\left|\Psi\right\rangle $ and $\left|\Phi_{1}\right\rangle $
can always be locally distinguished, the state of the preceding pair
is now determined using LOCC. 
\begin{itemize}
\item (a1.1) If the outcome is $\Psi$, it follows that all pairs $\left\{ i\right\} $
for $i\leq\left(i_{\text{mid}}-2\right)$ are in $\left|\Psi\right\rangle $,
 implying distillation of $\left(i_{\text{mid}}-2\right)$ pairs.
However, the original problem stays the same, for the change point
and mutation are yet to be found, but now it involves $\left(n-i_{\text{mid}}\right)$
pairs. 
\item (a1.2) If the outcome is $\Phi_{1}$, it follows that the mutation
is $\left|\Phi_{1}\right\rangle $ and all pairs from $\left(i_{\text{mid}}+1\right)$
to $n$ are in $\left|\Phi_{1}\right\rangle $. So this outcome leads
to distillation of $\left(n-i_{\text{mid}}\right)$ pairs. But, more
importantly, the original problem now reduces to the two-state problem
with orthogonal states $\left|\Psi\right\rangle $ and $\left|\Phi_{1}\right\rangle $
involving $\left(i_{\text{mid}}-2\right)$ pairs from $1$ to $\left(i_{\text{mid}}-2\right)$,
and thus can be efficiently solved. 
\end{itemize}
(a2) If the outcome is $01/10$, then the state was either $\left|\Phi_{2}\right\rangle $
or $\left|\Phi_{3}\right\rangle $, so one of them is the mutation.
Also, the next pair must be $\left|\Phi_{2}\right\rangle $ or $\left|\Phi_{3}\right\rangle $.
An LOCC measurement now identifies its state, and hence the mutation.
If the outcome is $\Phi_{2}$ (or $\Phi_{3}$), then all pairs from
$\left(i_{\text{mid}}+2\right)$ to $n$ must be in $\left|\Phi_{2}\right\rangle $
(or $\left|\Phi_{3}\right\rangle $); thus $\left(n-i_{\text{mid}}-1\right)$
entangled pairs are distilled. Therefore, as in (a1.2), the original
problem reduces to the two-state problem with orthogonal states $\left|\Psi\right\rangle $
and $\left|\Phi_{2}\right\rangle $ (or $\left|\Phi_{3}\right\rangle $)
with $\left(i_{\text{mid}}-1\right)$ pairs and can be efficiently
solved. 

So we see that in all cases except (a1.1), an efficient LOCC protocol
exists because the problem reduces to the two-state problem with orthogonal
states after the first two measurements. 

Now consider (a1.1). By following the same approach, it is sometimes
possible to identify both the mutation and the change point and distill
some more entanglement, but sometimes it is not. For example, if no
change occurs, the protocol fails to lead us to the correct conclusion.
Also, if the change happens at the last pair, the protocol could never
identify the mutation, though sometimes it could identify it as the
change point provided the mutation is $\left|\Phi_{2}\right\rangle $
or $\left|\Phi_{3}\right\rangle $. It is important to note that even
when the protocol fails to identify the mutation and/or the change
point, one could still distill entanglement. 

The quantum change point problem considers a faulty quantum source,
supposed to emit particles in a fixed state, but switches to emitting
a different state at some point. The task is to identify where the
change happened by measuring a sequence of emitted states. In this
paper, we studied this problem and a natural relaxation within the
paradigm of LOCC. Our study is motivated by entanglement-sharing protocols
that typically involve a source emitting entangled pairs in a fixed
state and are distributed to distant observers via quantum channels.
The purity of the shared entangled states is paramount as they are
resources for quantum information processing tasks. However, if the
source is faulty and switches to a different entangled state during
emission, the shared entangled states cannot be used because their
identities remain unknown. Clearly, this holds even in the absence
of decoherence effects and noisy channels. In other words, the quantum
change point problem in a LOCC setup has nontrivial consequences in
the context of quantum information. 

We showed that detecting the change point using LOCC frees up pure
entangled states akin to entanglement distillation. In particular,
if the default state and the mutated state are mutually orthogonal,
there exists an efficient LOCC protocol that detects the change point
and also distills a sufficiently large number $\approx n-\log_{2}n$
of free entangled pairs from a sequence of $n$ pairs. However, if
the default and the mutated states are nonorthogonal, then it may
not be possible to achieve both objectives. Our analysis showed that
one could nevertheless distill pretty good entanglement on average. 

We also considered a relaxation of the change point problem in which
the mutated state is unknown but belongs to a known set of entangled
states. We showed that change point detection and entanglement distillation
are connected to the LOCC distinguishability of the default and all
possible mutations: if they are LOCC distinguishable, the problem
reduces to the two-state scenario; otherwise, a desirable solution
may not be achievable. We discussed the latter in detail by considering
the set of four Bell states, one chosen as default and the rest as
possible mutations. Because the four Bell states are LOCC indistinguishable,
the two-state strategy that relied on identifying any pair by LOCC
is no longer applicable. We presented a modified protocol and found
that while distillation is always possible, and in most cases efficient,
there are instances where it fails to identify both the change point
and the mutation. However, a better protocol might be able to resolve
this issue. 

The results discussed in this paper can be straightforwardly extended
to a multipartite setting. The analysis will not change in the two-state
formulation with orthogonal states because any two multipartite orthogonal
states can also be locally distinguished. It will also not change
if one considers the variation of the two-state problem with locally
distinguishable states. 

There are a few open questions, nevertheless. The LOCC protocols discussed
in this paper appear pretty good but may not be optimal, and better
protocols could exist. Another interesting question arises in the
variation of the two-state problem: Can we find an example of a set
containing the default and all possible mutations, which is locally
indistinguishable, yet one can always identify the mutated state,
the change point, and, in addition, distill entanglement? 

There are several other related questions of interest that one might
look at. Following the standard change point problem, we considered
the situation where change occurs, if at all, only once. This could
be relaxed to allow the possibility of multiple change points. Another
way of approaching the LOCC change point problem is to consider the
so-called online detection method. Here, we have considered an offline
detection method, where the protocol requires quantum memory for practical
implementation--large enough for coherent storage and preservation
of the entangled pairs. This is not unique to this problem, but appears
in any entanglement distillation scenario. In order to address this,
one could measure entangled pairs as and when they are received. This
would be sub-optimal, of course, as many more entangled pairs would
be consumed before the change point is identified, and far less entangled
pairs would therefore be distilled. However, by computing the average
number of distilled entangled pairs, one could obtain a fair indication
of the efficiency of this method. But, this analysis is beyond the
scope of the present paper. 

Another problem for future research would be the noisy scenario in
which the quantum channels distributing the entangled pairs are noisy.
This problem has challenges of its own, as the resulting entangled
pairs will be mixed, and unambiguous identification of the change
point will not be possible. However, it would be interesting to know
whether efficient distillation protocols exist in this case. 
\begin{acknowledgments}
A.B. thanks Shayeef Murshid for useful discussions. 
\end{acknowledgments}

\section*{appendix: Two-state problem with orthogonal states}

\textbf{\emph{LOCC protocol}}

First, Alice and Bob determine the state of the pair at the midpoint
of the sequence $i_{\text{mid}}=\left\lfloor \frac{n}{2}\right\rfloor +1$
by means of LOCC that distinguishes between $\left|\Psi\right\rangle $
and $\left|\Phi\right\rangle $. If the outcome is $\Psi$, they conclude
that all (undisturbed) pairs on its left (the ones that came before)
are in $\left|\Psi\right\rangle $, and the change point lies on the
right of $i_{\text{mid}}$; however, if the outcome is $\Phi$, then
all (undisturbed) pairs on its right are in state $\left|\Phi\right\rangle $
and the change point is on the left of $i_{\text{mid}}$ or at $i_{\text{mid}}$.
Note that the $\Psi$ outcome distills $\left(i_{\text{mid}}-1\right)$
pairs {[}$1$ to $\left(i_{\text{mid}}-1\right)${]}, each in state
$\left|\Psi\right\rangle $; alternatively, the $\Phi$ outcome distills
$\left(n-i_{\text{mid}}\right)$ pairs {[}$\left(i_{\text{mid}}+1\right)$
to $n$ {]} in state $\left|\Phi\right\rangle $. \\
\\
If the first outcome is $\Psi$, they will work with $\left(n-i_{\text{mid}}\right)$
pairs on the right of $i_{\text{mid}}$, whereas, for the $\Phi$
outcome, they will work with $\left(i_{\text{mid}}-1\right)$ pairs
on the left of $i_{\text{mid}}$. Suppose the first outcome is $\Psi$.
They will now measure the pair $i_{\text{mid}}^{\prime}$, ``midway''
between $\left(i_{\text{mid}}+1\right)$ and $n$, to determine its
state. 
\begin{itemize}
\item If the outcome is $\Psi$, then it implies that the change point is
on the right of $i_{\text{mid}}^{\prime}$ and that all pairs from
$\left(i_{\text{mid}}+1\right)$ to $\left(i_{\text{mid}}^{\prime}-1\right)$
are in state $\left|\Psi\right\rangle $, and these are the pairs
they distill. 
\item However, if it is $\Phi$, they conclude that the change must have
occurred either between $i_{\text{mid}}$ and $i_{\text{mid}}^{\prime}$
or at $i_{\text{mid}}^{\prime}$; in this case, they also distill
a few more pairs, namely, those from $\left(k_{\text{mid}}^{\prime}+1\right)$
to $n$, all in the state $\left|\Phi\right\rangle $. 
\end{itemize}
On the other hand, if the first outcome is $\Phi$, they will measure
the pair $i_{\text{mid}}^{\prime\prime}$ at the ``mid-point'' between
$1$ and $\left(i_{\text{mid}}-1\right)$ and determine its state.
If this turns out to be $\left|\Psi\right\rangle $, the change point
lies between $i_{\text{mid}}^{\prime\prime}$ and $i_{\text{mid}}$
or at $i_{\text{mid}}$; however, if it is $\left|\Phi\right\rangle $
then the change point is on the left of $i_{\text{mid}}^{\prime\prime}$
or at $i_{\text{mid}}^{\prime\prime}$. In either case, they distill
a few more pairs.

This procedure is repeated, where the outcome at any stage singles
out a subset of pairs to explore further and leaves out its complement
(contains identified pairs); in particular, the $\Psi$ outcome pushes
them to the right and $\Phi$ to the left of the sequence. Observe
that after every measurement, the number of pairs they need to focus
on reduces roughly by a factor of two. It follows that after $\left(\left\lfloor \log_{2}\left(n+1\right)\right\rfloor -1\right)$
measurements, they will be left with a single pair, or, at most two--if
left with one, determining its state identifies the change point (which
could be either that or the next); if left with two pairs, sometimes
measuring one is enough, else they need to determine the states of
both. \\
\textbf{\emph{Required number of measurements}}

First note that after the first measurement we have used one pair
and are left with two subsequences to explore depending on the measurement
outcome. The larger of the two subsequences is always of length $\left\lfloor \frac{n}{2}\right\rfloor $.
To encounter the worst case, we must consistently drift towards the
larger subarray for the next measurement, continuing this process
iteratively.  From this, we can easily arrive at the recurrence relation
for the number of iterations in the worst-case case scenario, 

\begin{alignat*}{1}
B_{n} & =B_{\left\lfloor \frac{n}{2}\right\rfloor }+1
\end{alignat*}
that can be solved exactly to give us the solution
\begin{alignat*}{1}
B_{n} & =\left\lfloor \log_{2}n\right\rfloor +1.
\end{alignat*}

Let us now look at the best case. Once again, after the first measurement,
which consumes one pair, we are left with one of the two subsequences
depending on the outcome. The smaller of the two is of length $\left\lfloor \frac{n-1}{2}\right\rfloor $.
For the best case to happen, we must always drift to the smaller of
the two subsequences for the next round of measurements. From this
we arrive at the recurrence relation for the number of iterations
\begin{alignat*}{1}
B_{n} & =B_{\left\lfloor \frac{n-1}{2}\right\rfloor }+1
\end{alignat*}
with the initial conditions $B_{1}=1$ and $B_{2}=1$. This can be
exactly solved to give us
\begin{alignat*}{1}
B_{n} & =\left\lfloor \log_{2}\left(n+1\right)\right\rfloor .
\end{alignat*}
 Observe that the required number of measurements in the two scenarios
differs, at most, by $1$, so the protocol will require either $\left\lfloor \log_{2}\left(n+1\right)\right\rfloor $
or $\left(\left\lfloor \log_{2}n\right\rfloor +1\right)$ measurements.\\
\textbf{\emph{Number of distilled pairs}}

Since every LOCC measurement consumes an entangled pair, the number
of residual (distilled) entangled pairs is either $\left(n-\left\lfloor \log_{2}\left(n+1\right)\right\rfloor \right)$
or $\left(n-\log_{2}\left\lfloor n\right\rfloor -1\right)$. In particular,
the residual pairs on the left of the change point are in $\left|\Psi\right\rangle $,
and those on the right are in $\left|\Phi\right\rangle $. However,
all residual pairs will be in $\left|\Phi\right\rangle $ if the change
happens in the first position or $\left|\Psi\right\rangle $ if there
is no change.

\newpage

\begin{thebibliography}{99}
\bibitem{Note-1} The key assumption is that $\left|\phi\right\rangle $
is known beforehand. This assumption could be relaxed as shown in
this paper.
\bibitem{Note-2} Entanglement distillation, after all, means a LOCC
procedure to obtain free entangled pairs from a collection of mixed
states that are distillable. The procedure depends entirely on the
ensemble being considered. As we will show, detecting the CP using
LOCC leads to free entangled pairs, so this ought to be treated as
bonafide distillation.
\bibitem{prior-probs} These probabilities do not make any difference
in the deterministic scenario where the default and the mutation can
be perfectly distinguished by LOCC.
\end{thebibliography}
\end{document}